\documentclass{article}
\usepackage{amsmath}
\usepackage{amsfonts}
\usepackage{amsthm}

\newcommand{\Section}[1]{\section{#1} \setcounter{equation}{0}}

\newcommand{\dt}{\frac{d}{dt}}
\newcommand{\drs}{\frac{\partial}{\partial r_*}}
\newcommand{\drss}{\frac{\partial^2}{\partial r_*^2}}
\newcommand{\Reals}{\mathbf R}

\newcommand{\dalembertian}{\Box}
\newcommand{\defin}{\equiv}
\newcommand{\starman}{\mathfrak{M}}
\newcommand{\hide}[1]{}

\renewenvironment{proof}{ Proof:}{\hfill $\Box$}
\newtheorem{theorem}{Theorem}

\newtheorem{lemma}[theorem]{Lemma}
\newtheorem{remark}[theorem]{Remark}
\newtheorem{definition}[theorem]{Definition}

\newcommand{\pushleft}{}

\title{Semilinear wave equations on the Schwarzschild manifold \newline I: Local Decay Estimates.}
\author{ P. Blue\\ \small Mathematics dept., Rutgers University\\ \small Piscataway, NJ 08854, USA\\ \and
A. Soffer\\ \small Mathematics dept., Institute for Advanced Studies\\ \small Princeton, NJ 08540, USA}
\date{17 October 2003}

\begin{document}
\maketitle

Appeared in {\em Advances in Differential Equations}, 8 (2003) 595-614. 

\begin{abstract} 
The semilinear wave equation on the (outer) Schwarzschild manifold is studied. We prove local decay estimates for general (non-radial) data, deriving a-priori Morawetz type inequalities.
\end{abstract}

\Section{Introduction}
The General Theory of Relativity has solutions for space-time of the black-hole type. The simplest example is the Schwarzschild solution. The analysis of the equations of General Relativity near a Schwarzschild solution motivates us to study the wave propagation in such a metric. In this work we analyze semi-linear wave equations in such a metric. Our main objective is to construct a good set of a-priori estimates, modelled after the Morawetz estimate in the flat case, and then use it to prove local decay and other estimates for the non-linear wave equation on the Schwarzschild manifold. Our methods are inspired in part by the previous work on the non-linear Schr\" odinger equation \cite{LabaSoffer}. We deal with general, non-radial initial conditions. 

The (external) Schwarzschild manifold is $(\Reals\times(2M,\infty)\times S^2)$ with the metric (written in polar co-ordinates)
\begin{equation}
ds^2 = (1-\frac{2M}{r})dt^2 - (1-\frac{2M}{r})^{-1}dr^2 -r^2(d\theta^2+\sin^2(\theta)d\phi^2)
\end{equation}

Scattering theory for the wave and Klein-Gordon equations on the Schwarzschild manifold was first studied by Dimock \cite{Dimock} and Dimock and Kay \cite{DimockKay}. Following these authors, we introduce the Regge-Wheeler tortoise co-ordinate
\begin{equation}
r_* = r +2M\log(\frac{r-2M}{2M})
\end{equation}
and treat $r$ as a function of $r_*$ for which
\begin{equation}
\frac{dr}{dr_*} = (1-\frac{2M}{r})
\end{equation}
Using this new co-ordinate the metric becomes
\begin{equation}
ds^2 = (1-\frac{2M}{r})dt^2 - (1-\frac{2M}{r})dr_*^2 - r^2(d\theta^2+\sin^2(\theta)d\phi^2)
\end{equation}
The associated d'Alembertian operator is
\begin{equation}
\dalembertian = (1-\frac{2M}{r})^{-1}(\frac{\partial^2}{\partial t^2}-r^{-2}\drs r^2\drs) -r^{-2}\Delta_{S^2}
\end{equation}
where $\Delta_{S^2}$ is the Laplacian on the unit sphere.

We wish to consider the non-linear wave equation
\begin{equation}
\dalembertian u = -\lambda|u|^{p-1}u
\end{equation}
which is equivalent to
\begin{equation}
\label{tildenonlineqn}
\frac{\partial^2}{\partial t^2} u + \tilde{H}_p u = 0
\end{equation}
where
\begin{eqnarray}
\tilde{H}_p &=& \tilde{H}+ \lambda(1-\frac{2M}{r})|u|^{p-1}\\
\tilde{H} &=& -r^{-2}\drs r^2\drs - (1-\frac{2M}{r})r^{-2}\Delta_{S^2} 
\end{eqnarray}

The existence and completeness of wave operators for the linear wave equation was proven in \cite{Dimock}. For the Klein-Gordon equation $\dalembertian u+m^2u=0$, existence of wave operators was proven in \cite{DimockKay} and asymptotic completeness in \cite{Bachelot}. In \cite{FroeseHislop} a Mourre estimate was obtained which was then used to recover the scattering results of \cite{Dimock} on a more general class of non-compact manifolds. 

As shown recently in \cite{YauSmoller} the local decay estimate in the presence of black holes is more subtle than the flat case. In particular for massive Dirac fields the $L^\infty$ norm decays generically as $t^{-\frac{5}{6}}$ rather than than $t^{-\frac{3}{2}}$ in the flat case. 

\cite{BachelotNicolas, Nicolas} have already proven partial results for the non-linear wave equation on more general manifolds than we consider. In particular, Nicolas \cite{Nicolas} obtained global existence and an outgoing radiation condition for the Klein-Gordon equation with an explicit non-linearity. In \cite{Shu}, the Cauchy problem for the Yang-Mills equation on the Schwarzschild manifold was studied.

In this paper we prove local decay for the non-linear wave equation on the Schwarzschild manifold, in the $L^2$ sense. We do not use radial symmetry of the initial conditions as in previous work \cite{LabaSoffer}; therefore, we do not place any conditions on the angular behaviour as in \cite{YauSmoller}. Our proof is based on showing the modified Morawetz estimate holds for the wave equation. It is expected that these estimates can then be used to derive other, more general a-priori estimates. This will be done elsewhere. 

In section \ref{secPrelim} we note some basic properties of the $r_*$ co-ordinate. In section \ref{secHeisenbergType}, we develop a formula relating the commutator of an operator with the Hamiltonian to the derivative of a certain inner product involving the operator. We use this to show conservation of energy. In section \ref{secChangeL2}, we use a unitary transform of $L^2$ to write the Hamiltonian as the sum of a constant coefficient second order radial derivative, angular derivatives, and a potential term. This allows us to write the energy in a more familiar form, which we use to bound the growth of the $L^2$ norm of the solutions to the non-linear wave equation. In section \ref{secMorawetzMult}, we introduce a Morawetz style radial multiplier. This multiplier is based on the one found in \cite{LabaSoffer, Lavine, MorawetzStrauss}, modified to handle the non-radial case. In section \ref{secMorawetzEst}, we prove an analogue of the Morawetz estimate and use this in section \ref{secLocalDecay} to prove local decay for (\ref{tildenonlineqn}). 

\Section{Preliminaries}
\label{secPrelim}
In this section we note some basic physical properties of the system. 

We treat $r$ as a function of $r_*$ and note that it is increasing and has the properties that as $r_*\rightarrow\infty, r\rightarrow\infty$, and as $r_*\rightarrow -\infty, r\rightarrow 2M$ and $1-\frac{2M}{r}\rightarrow 0$ (cf \cite{LabaSoffer}). 

There is a conserved energy defined by
\begin{equation}
E(u) = \int_{S^2}\int_{-\infty}^{\infty} (\bar{u}\tilde{H}u +\frac{2\lambda |u|^{p+1}}{(p+1)r^{p-1}} + |\dot{u}|^2)r^2dr_* d\omega
\end{equation}
This is proven in theorem \ref{EConsGen}. 

It is assumed throughout that $\tilde{H}$ is defined on its natural domain as a self-adjoint operator (cf \cite{Bachelot}). 

In the case $\lambda >0$, we derive all estimates on the interval of existence of the solutions. For appropriate choice of $p$ ($p<5$) global existence may follow from our estimates for large initial data. For small initial data this follows by standard fixed point methods. This will be discussed elsewhere. 

\Section{Heisenberg relation for the wave equation}
\label{secHeisenbergType}
Motivated by the Heisenberg equation for the Schr\" odinger equation, we derive an equation relating the time derivative of an operator to its commutator with the Hamiltonian, $\tilde{H}_p$. The formula holds for a general wave equation, and in particular will continue to hold once we have changed co-ordinates in section \ref{secChangeL2}. When the commutator is zero we derive a conserved quantity and when the commutator is positive we can derive time decay. We first use this to derive energy conservation, and later use a different operator to prove an analogue of the Morawetz estimate. 

\begin{theorem}
\label{ComRelthm}
For a time independent operator $A$ and a solution $u$ to the nonlinear wave equation, $\ddot{u}+\tilde{H}_p u=0$, such that $u$, $u^p$, and $ \tilde{H} u$ are in the domain of $A$, $u$ and $Au$ are in the domain of $\tilde{H}$, and $u^p$ is in $L^2$, 
\begin{equation}
\label{ComRel}
\dt(\langle u,A\dot{u}\rangle - \langle\dot{u},Au\rangle) = \langle u,[\tilde{H}_p,A]u\rangle
\end{equation}
\end{theorem}
\begin{proof}
Since $u$ is a solution of the nonlinear wave equation $\frac{\partial^2}{\partial t^2}u=-\tilde{H}_p u$,
\begin{eqnarray}
\dt(\langle u,A\dot{u}\rangle-\langle\dot{u},Au\rangle) &=& \langle\dot{u},A\dot{u}\rangle+\langle u,A\ddot{u}\rangle-\langle\ddot{u},Au\rangle-\langle\dot{u},A\dot{u}\rangle \nonumber\\
&=& -\langle u,A\tilde{H}_p u\rangle+\langle \tilde{H}_p u,Au\rangle \nonumber\\
&=& \langle u,[\tilde{H}_p,A]u\rangle \nonumber
\end{eqnarray}
\end{proof}

\begin{remark}
Frequently it is useful to use a time dependent operator $A$; in this case there are the following two additional terms on the right hand side of equation (\ref{ComRel}). 
\begin{equation}
\langle u,(\dt A)\dot{u}\rangle -\langle\dot{u},(\dt A)u\rangle
\end{equation}
\end{remark}

\begin{remark}
We expect that equation \ref{ComRel} holds for a significantly larger class of $u$. 
\end{remark}

We now use theorem \ref{ComRelthm} to prove conservation of energy. 

\begin{theorem}
\label{EConsGen}
The non-linear wave equation has a conserved quantity $\|u\|_{\mathcal{H}}^2$ which we call the energy. This holds at least for $H^2(\Reals^2\times S^2)$. 
\begin{equation}
\|u\|_{\mathcal{H}}^2 \defin \langle\dot{u},\dot{u}\rangle +\langle u,\tilde{H}u\rangle +\frac{2}{p+1}\langle u,\lambda |u|^{p-1}u\rangle
\end{equation}
\end{theorem}
\begin{proof}
If $u$ is a solution to the non-linear wave equation
\begin{equation}
\dt(\langle\dot{u},\dot{u}\rangle + \langle u,\tilde{H}u\rangle + \langle u,\lambda |u|^{p-1}u\rangle) \pushleft
\end{equation}
\begin{eqnarray}
&=&\dt(\langle\dot{u},\dot{u}\rangle + \langle u,\tilde{H}_pu\rangle)\nonumber\\
&=&\dt(\langle\dot{u},\dt u\rangle-\langle u,\dt \dot{u}\rangle)\nonumber\\
&=&-\langle u,[\tilde{H}_p,\dt]u\rangle\nonumber\\
&=&\langle u,\lambda \dt(|u|^{p-1})u\rangle \nonumber\\
&=&\frac{p-1}{p+1} \dt\langle u,\lambda |u|^{p-1}u\rangle\nonumber
\end{eqnarray} 
\end{proof} 

Our method is to use the Heisenberg relation for the wave equation to provide conserved quantities and estimates for the wave equation. In the case of the energy, we were fortunate that the commutator term could be written as an exact time derivative which gave a conserved quantity. Typically the commutator term will be more difficult to deal with; however, if it has a sign we will be able to derive an estimate, as in the following theorem. This will be our general technique and our main concern will be to provide bounds on the commutator. The essence of the proof is to integrate the Heisenberg relation for the wave equation. 

\begin{theorem}
For a solution $u$ to the non-linear wave equation, and a time independent operator $A$ such that the commutator $[\tilde{H}_p,A]\geq B$, then 
\begin{equation}
(\langle u,A\dot{u}\rangle-\langle\dot{u},Au\rangle)\mid_0^T \geq \int_0^T\langle u,Bu\rangle dt 
\end{equation}
\end{theorem}
\begin{proof}
\begin{eqnarray}
(\langle u,A\dot{u}\rangle-\langle\dot{u},Au\rangle)\mid_0^T &=& \int_0^T \dt(\langle u,A\dot{u}\rangle-\langle\dot{u},Au\rangle)dt\nonumber\\
&=& \int_0^T\langle u,[\tilde{H}_p,A]u\rangle dt\nonumber\\
&\geq&\int_0^T \langle u,Bu\rangle dt
\end{eqnarray}
\end{proof}

\Section{Change of $L^2$ space}
\label{secChangeL2}
In order to further simplify the problem, we will define a new manifold, $\starman$, and an isometry of $L^2$ functions on our original manifold to $L^2$ functions on this new manifold. Unless otherwise specified all inner products, $\langle\circ,\circ\rangle$, and norms, $\|\circ\|$, will be assumed to be in this new $L^2$ space. This change of variables is inspired by \cite{LabaSoffer} and introduces an effective potential $V$ which will appear in the energy. 

\begin{definition}
\begin{eqnarray}
\starman &\defin& (\Reals\times S^2,ds^2=dr_*^2+ds_{S^2}^2)\\
d^3\mu &\defin& dr_*d^2\omega \\
U &:& L^2(\starman ,d^3\mu) \rightarrow L^2(\Reals\times S^2,r^2dr_*d\omega)\\
u(r_*,\theta,\phi)&\rightarrow& \frac{u(r_*,\theta,\phi)}{r(r_*)}
\end{eqnarray}
\end{definition}

\begin{theorem}
\label{UIsometry}
$U$ is an isometry of $L^2$ spaces. $\tilde{H}_p$ is mapped to $H_p$ where, 
\begin{eqnarray}
V &\defin& \frac{2M}{r^3}(1-\frac{2M}{r}) \\
H &\defin& -\drss + V -\frac{1}{r^2}(1-\frac{2M}{r})\Delta_{S^2} \\
H_p &\defin& H + \lambda (1-\frac{2M}{r}) r^{1-p}|u|^{p-1}u
\end{eqnarray}
and the non-linear wave equation becomes
\begin{equation}
\label{TransWE}
(\frac{\partial^2}{\partial t^2} +H_p) u = 0
\end{equation}
\end{theorem}
\begin{proof}
The original measure is $d^3\mu=r^2 dr_*d^2\omega$, which differs from the new measure exactly by a factor of $r^2$, so division of the functions by $r$ will correct for this in $L^2$ norm. Thus $U$ preserves norm. It is also one to one and onto, and hence an isometry. 
We now calculate the image of the Hamiltonian, by calculating its action on $U^{-1}u$ for some $u\in L^2(\Reals\times S^2,r^2 dr d\omega)$, 
\begin{eqnarray}
U^{-1}\tilde{H}_pU U^{-1}u &=& U^{-1}(-r^{-2}\drs r^2\drs -r^{-2}(1-\frac{2M}{r})\Delta_{S^2})u + U^{-1}\lambda(1-\frac{2M}{r}) |u|^{p-1}u \nonumber \\
&=& (-r^{-1}\drs r^2\drs r^{-1}- r^{-1}(1-\frac{2M}{r})\Delta_{S^2}r^{-1})U^{-1}u + \lambda(1-\frac{2M}{r}) |u|^{p-1} U^{-1}u  \nonumber\\
&=& (-r^{-1}\drs(r\drs-(1-\frac{2M}{r})) -r^{-2}(1-\frac{2M}{r})\Delta_{S^2})U^{-1}u \nonumber \\
&& + \lambda(1-\frac{2M}{r}) r^{1-p}|U^{-1}u|^{p-1} U^{-1}u \nonumber\\
&=& (-\drss +\frac{2M}{r^3}(1-\frac{2M}{r}) -r^{-2}(1-\frac{2M}{r})\Delta_{S^2}) U^{-1}u + \lambda(1-\frac{2M}{r}) r^{1-p}|U^{-1}u|^{p-1} U^{-1}u \nonumber
\end{eqnarray}
This gives the image of the Hamiltonian under $U$. Since $U$ commutes with the time derivative, the wave equation acting on $L^2(\starman,d^3\mu)$ becomes equation (\ref{TransWE}). 
\end{proof}

\begin{remark}
If $u\in L^2(\starman,d^3\mu)$ and $u$ is a pure angular momentum state, that is the product of a radial function with a solution of the spherical wave equation, then $-\Delta_{S^2}u=l(l+1)u$, and the linear part of the Hamiltonian acts by
\begin{equation}
H u= (-\drss + V + V_l)u
\end{equation}
where
\begin{equation}
V_l \defin \frac{l(l+1)}{r^2}(1-\frac{2M}{r})
\end{equation}
The spherical harmonics form a complete orthonormal system, and we can define projection operators $P_l$ onto the states of pure angular momentum. Using these, we can rewrite the linear part of the Hamiltonian as
\begin{equation}
H = -\drss + V + \sum_{l=0}^{\infty}V_l P_l
\end{equation}
\end{remark}

\begin{remark}
It is now possible to show that $H$ is self-adjoint by W\" ust's theorem and that $C^{2}(\starman)$ is a core. Dimock \cite{Dimock} proves self-adjointness via a decomposition to spherical harmonics with a slightly different core. 
\end{remark}

\begin{remark}
We will sometimes write $u'$ for $\drs u$. This should not be confused with the gradient which we never need to compute. 
\end{remark}

The change of co-ordinates allows us to rewrite the energy in a more familiar form, as a sum of terms coming from the time derivative, the radial derivative, the angular derivatives, the potential, and the non-linear terms. 

\begin{theorem}
\label{ECons}
The energy in the $\starman$ co-ordinates is
\begin{eqnarray}
\|u\|_{\mathcal{H}}^2 &=& \langle\dot{u},\dot{u}\rangle + \langle u',u'\rangle +\langle u,Vu\rangle + \sum_{l=0}^{\infty}\langle P_lu,V_lP_lu\rangle + \frac{2}{p+1}\langle u,\lambda r^{1-p}(1-\frac{2M}{r}) |u|^{p-1} u\rangle
\end{eqnarray}
This acts as a metric on the space $\mathcal{H} \defin \{u\in L^2\mid \|u\|_{\mathcal{H}}<\infty\}$.
\end{theorem}
\begin{proof}
This is a simple restatement of theorem \ref{EConsGen}. The metric properties follow from the metric properties of $L^2$ and the positivity of the potentials $V$ and $V_l$ and of the non-linearity. 
\end{proof}

Finally we use the energy to provide a bound on the growth of the $L^2$ norm. 

\begin{theorem}
\label{L2Growth}
If $u$ is a real valued solution of the non-linear wave equation, and $\|u(t)\|_{L^2}$ is the $L^2$ norm of $u$ at time $t$, 
\begin{equation}
\|u(t)\|_{L^2} \leq t \|u\|_{\mathcal{H}} + \|u(0)\|_{L^2}
\end{equation}
(Recall that $\|u\|_{\mathcal{H}}$ is constant.)
\end{theorem}
\begin{proof}
\begin{eqnarray}
\dt \|u\|_{L^2}^2 &=& \dt\langle u,u\rangle \nonumber\\
&=& 2\langle u,\dot{u}\rangle \nonumber\\
2\|u\|_{L^2} \dt\|u\|_{L^2}&\leq& 2\|u\|_{L^2} \|\dot{u}\|_{L^2} \nonumber\\
\dt \|u\|_{L^2} &\leq& \|u\|_{\mathcal{H}} \nonumber
\end{eqnarray}
\end{proof}

\Section{The Morawetz type multiplier}
\label{secMorawetzMult}
We now introduce a first order differential operator which is one possible analogue of the Morawetz multiplier. Because our change of variables has introduced a radial parameter extending from negative infinite to positive infinite, it is necessary to choose a radius which corresponds to the origin of the standard Morawetz multiplier. We choose the radius which maximises the angular potential $V_l$. We call this the centrifugal tortoise origin since in the tortoise co-ordinates the angular potential is repulsive about this radius. We then go onto prove some basic properties following from the definition. 

\begin{definition}
The centrifugal origin $\alpha$, centrifugal tortoise origin $\alpha_*$, and centrifugal tortoise radius $\rho_*$ are defined by
\begin{eqnarray}
\alpha &\defin& 3M \\
\alpha_* &\defin& 3M + 2M\log(\frac{1}{2}) \\
\rho_* &\defin& r_* -\alpha_* 
\end{eqnarray}
\end{definition}

\begin{remark}
Later calculations will show that the angular potential $V_l$ has a maximum at $r=\alpha$ which corresponds to $r_*=\alpha_*$. Those calculations will be simplified by using $\rho_*$. 
\end{remark}

In the following definition we take $\sigma\geq\frac{1}{2}$ so that $g_\sigma$ is bounded. 

\begin{definition}
Given $\sigma\in(\frac{1}{2},\frac{3}{2})$, the Morawetz type multiplier $\gamma_\sigma$ is defined by
\begin{eqnarray}
g_\sigma(r_*) &=& \int_0^{\frac{r_*-\alpha_*}{2M}} (1+\tau^2)^{-\sigma} d\tau \\
\gamma_\sigma &=& -\frac{i}{2}(g_\sigma(r_*)\drs + \drs g_\sigma(r_*)) \\
D(H,\gamma_\sigma) &\defin& \{ u| u,Hu\in D(\gamma_\sigma), u,\gamma u\in D(H)\}\\
D(H_p,\gamma_\sigma) &\defin& \{ u| u,H_pu\in D(\gamma_\sigma), u,\gamma u\in D(H), u^p\in L^2\}
\end{eqnarray}
where for an operator $A$, $D(A)$ is its domain. 

In all cases we will deal with a fixed value of $\sigma$ and will write $\gamma=\gamma_\sigma$ and $g=g_\sigma$.
\end{definition}

\begin{theorem}
\label{P1Mm}
If $u$ is a real valued solution of the non-linear wave equation in $H^1(\starman)$ and $\sigma\in(\frac{1}{2},\frac{3}{2})$, then $\langle u,\gamma_\sigma u\rangle=0$.
\end{theorem}
\begin{proof}
\begin{eqnarray}
\langle u,\gamma u\rangle &=& \langle u, -\frac{i}{2}(g\drs +\drs g) u\rangle \nonumber\\
&=& -\frac{1}{2}(\langle u, ig\drs u\rangle +\langle u, i \drs gu\rangle)\nonumber\\
&=& -\frac{1}{2}(\langle u,ig\drs u\rangle + \langle g i\drs u,u\rangle)\nonumber\\
&=& -\Re \langle u, ig\drs u\rangle \nonumber\\
&=& 0 \nonumber
\end{eqnarray}
since $g, u$, and its derivatives are all real valued. 
\end{proof}

\begin{theorem}
\label{P2Mm}
If $u$ is a real valued solution of the non-linear wave equation in $H^1(\starman)$ and $\sigma\in(\frac{1}{2},\frac{3}{2})$, then $\exists C_\sigma,D$:
\begin{equation}
\|\gamma u\|_{L^2}\leq C_\sigma\|u\|_{\mathcal{H}}+D\|(1+(\frac{r_*-\alpha_*}{2M})^2)^{-\sigma} u\|_{L^2}
\end{equation}
\end{theorem}
\begin{proof}
We first note that the integrand in the definition of $g$ is continuous, even, and decays like $\tau^{-2\sigma}$ with $\sigma>\frac{1}{2}$. Thus $g$ is bounded by a constant $C_\sigma$.
\begin{eqnarray}
\|\gamma u\| &=& \|g u' + \frac{1}{2}g'u\|\nonumber\\
&\leq& \|gu'\|+\frac{1}{2}\|g' u\| \nonumber\\
&\leq& C_\sigma \|u\|_{\mathcal{H}} + D\|(1+(\frac{r_*-\alpha_*}{2M})^2)^{-\sigma} u\|\nonumber
\end{eqnarray}
\end{proof}

\Section{The Morawetz type estimate}
\label{secMorawetzEst}
Our goal in this section is to prove theorem \ref{Morawetzthm} which bounds the commutator of the Morawetz type multiplier and the Hamiltonian from below by local decay terms. The commutator is linear so we can compute the effect of each term in the Hamiltonian separately. Estimates of the second radial derivative contributions provide the main term. The potential $V$ contributes terms of the wrong sign, but its contributions are dominated by the contributions from the second radial derivative. The angular derivative terms provide positive contributions which can be ignored, since they only strengthen the result. The non-linear term provides a non-linear local decay result. Since we choose the non-linearity to be repulsive, $\lambda\geq 0$, this will further strengthen theorem \ref{Morawetzthm}. 

We first prove the commutator of the angular derivative and our multiplier is positive. Since positive terms in the commutator will only strengthen the result desired in theorem \ref{Morawetzthm}, we can ignore the angular derivatives after proving this lemma. 

\begin{lemma}
\label{MorawetzAngular}
For $\sigma\in(\frac{1}{2},\frac{3}{2})$,
\begin{equation}
i[-\frac{1}{r^2}(1-\frac{2M}{r})\Delta_{S^2},\gamma] \geq 0
\end{equation}
More precisely, the commutator is a positive operator. 
\end{lemma}
\begin{proof}
Since the projection onto angular momentum subspaces commutes with the radial derivative and radial functions, it is sufficient to work on the angular momentum subspaces when proving the positivity of the commutator involving the spherical Laplacian. For such functions, the spherical Laplacian acts as the potential $V_l$. 
\begin{eqnarray}
V_l &=& \frac{l(l+1)}{r^2} \left(1-\frac{2M}{r}\right) \\
i[V_l,\gamma] &=& -g(r_*)\drs \left[\frac{l(l+1)}{r^2}\left(1-\frac{2M}{r}\right)\right]\nonumber\\
&=&-g(r_*)\frac{l(l+1)}{r^2}\left[\frac{-2}{r}\left(1-\frac{2M}{r}\right)+\frac{2M}{r^2}\right]\left(1-\frac{2M}{r}\right) \nonumber\\
&=& -g(r_*)\left[\frac{3M}{r}-1\right]\frac{2l(l+1)}{r^3}\left(1-\frac{2M}{r}\right)\nonumber
\end{eqnarray}
All these terms are positive except for $-g(r_*)$ and $\frac{3M}{r}-1$, the product of which is non-negative by the choice $\alpha=3M$. The resulting product is non-negative which proves the desired result. 
\end{proof}

We now prove an identity relating the commutator of the second radial derivative and the Morawetz type multiplier to a second order operator involving the function $g$ which was used in the definition of the Morawetz multiplier. We will then use this identity to bound commutator from below. 

\begin{lemma}
\label{MorawetzdrssAsDerivatives}
For $\sigma\in(\frac{1}{2},\frac{3}{2})$,
\begin{equation}
i[-\drss,\gamma] = -2\drs g'\drs-\frac{1}{2}g''' 
\end{equation}
\end{lemma}
\begin{proof}
\begin{eqnarray}
i[-\drss,\gamma] &=& \frac{-1}{2}\left(\drss(g\drs+\drs g)-(g\drs+\drs g)\drss\right)\nonumber\\
&=&\frac{-1}{2}\left(\drss g\drs +\frac{\partial^3}{\partial r_*^3}g-g\frac{\partial^3}{\partial r_*^3}-\drs g\drss\right)\nonumber\\
-2i[-\drss,\gamma]&=&\drs g'\drs +\drs g\drss+\drss g'+\drss g\drs-g\frac{\partial^3}{\partial r_*^3}-\drs g\drss \nonumber\\
&=& \drs g' \drs +\drs g'' +\drs g'\drs+\drs g'\drs+\drs g\drss-g\frac{\partial^3}{\partial r_*^3} \nonumber\\
&=& 3\drs g'\drs + g'''+g''\drs+g'\drss+g\frac{\partial^3}{\partial r_*^3} -g\frac{\partial^3}{\partial r_*^3} \nonumber\\
&=&3\drs g\drs+g'''+g''\drs+g'\drss \nonumber\\
i[-\drss,\gamma] &=& -2\drs g'\drs-\frac{1}{2}g''' \label{drssgammacommutatorterm}
\end{eqnarray}
\end{proof}

Having proved the previous identity, we use it bound the commutator of the second radial derivative and the Morawetz multiplier from below. The proof of the following lemma is based on direct, but sometimes lengthy, computation. The expectation value of the term in equation \ref{drssgammacommutatorterm} can be written as the sum of two expectation values. One is a differential operator and the other is a multiplication operator. The expectation value of the multiplication operator is estimated directly from known properties of $g$ and its derivatives. The remaining differential operator can be rewritten in the form $\partial_i c^{ij}\partial_j$, with $c^{ij}$ a (3x3) positive definite matrix. 


\begin{lemma}
For $\sigma\in(\frac{1}{2},\frac{3}{2})$ and $u\in D(H,\gamma)$, 
\begin{equation}
\label{drssbound}
\langle u,i[-\drss,\gamma]u\rangle \geq \langle u, \frac{\sigma}{(1+(\frac{\rho_*}{2M})^2)^{\sigma+2}}\frac{1}{(2M)^3}\left[5+(3-2\sigma)(\frac{\rho_*}{2M})^2\right] u\rangle
\end{equation}
\end{lemma}
\begin{proof}
We will convert the inner product into an integral which we break into two pieces. One of the pieces is positive and the other provides the bound of the form we want. Note that we use $\defin$ when defining a new term or function. 
\begin{eqnarray}
\langle u,i[-\drss,\gamma]u\rangle &=& \int\bar{u}(-2\drs g'\drs -\frac{1}{2}g''')u dr_* d\omega\nonumber\\
&=& \int ( \rho_*^{-1}\bar{u} \rho_*^{-1}(-2\drs g'\drs -\frac{1}{2}g''')\rho_* \rho_*^{-1}u ) \rho_*^2 dr_* d\omega\nonumber\\
&=& \int_\starman ( (\rho_*^{-1}u) \rho_*^{-1}(-2\drs g'-2 \drs g' \rho_*\drs -\frac{\rho_*}{2}g''') (\rho_*^{-1}u) ) \rho_*^2dr_*  d\omega\nonumber\\
&=& \int_\starman ( (\rho_*^{-1}\bar{u}) \rho_*^{-1}(-2g''-2g'\drs-2g''\rho_*\drs-2g'\drs-2g'\rho_*\drss-\frac{\rho_*}{2}g''') (\rho_*^{-1}u) ) \nonumber\\
&& \rho_*^2dr_*  d\omega\nonumber\\
&=& \int_\starman ( (\rho_*^{-1}\bar{u}) (-2\rho_*^{-1}g''-4\rho_*^{-1}g'\drs-2g''\drs-2g'\drss-\frac{g'''}{2}) (\rho_*^{-1}u) ) \rho_*^2dr_*  d\omega\nonumber\\
&=& \int_\starman ( (\rho_*^{-1}\bar{u}) (-2\rho_*^{-1}g'' -4\rho_*^{-1}g'\drs-2\drs g'\drs -\frac{g'''}{2}) (\rho_*^{-1}u) ) \rho_*^2dr_* d\omega \nonumber\\
&=& \int_{S^2} I_1 + I_2 d\omega \\
I_1 &\defin& \int_{-\infty}^{\infty}  ( (\rho_*^{-1}\bar{u}) (-4\rho_*^{-1}g'\drs-2\drs g'\drs) (\rho_*^{-1}u) ) \rho_*^2 dr_* \\
I_2 &\defin& \int_{-\infty}^{\infty} (\rho_*^{-1}\bar{u}) (-2\rho_*^{-1}g''-\frac{g'''}{2}) (\rho_*^{-1}u) \rho_*^2dr_* \nonumber\\
&=& \int_{-\infty}^{\infty} \bar{u} (-2\rho_*^{-1}g''-\frac{g'''}{2}) u dr_* 
\end{eqnarray}

We will first deal with the $I_1$ term. We will use $\rho_*$ as a variable instead of a function, then following \cite{LabaSoffer} break $I_1$ into two integrals on $\Reals^+$ of the same form, and treat these as integrals in $\Reals^3$ with $\rho_*$ acting as a radial parameter. We define some new functions and a new operator, and use these to rewrite $I_1$. 
\begin{eqnarray}
h(\rho_*) &\defin& g(r_*) = g(\rho_*+\alpha_*) \nonumber\\
h'(\rho_*) &=& h'(-\rho_*) \nonumber\\
f(\rho_*) &\defin& \rho_*^{-1}u(\rho_*+\alpha_*,\omega) = \rho_*^{-1}u(r_*,\omega) \in L^2(\Reals,\rho_*^2d\rho_*) \nonumber\\
L &\defin& -2\left(\frac{2h'(\rho_*)}{\rho_*}\frac{\partial}{\partial \rho_*} +\frac{\partial}{\partial \rho_*} h'(\rho_*)\frac{\partial}{\partial \rho_*} \right) \nonumber\\ 
\frac{-1}{2}L &=& \left(\frac{2h'(\rho_*)}{\rho_*} + h''(\rho_*) + h'(\rho_*)\frac{\partial}{\partial \rho_*}\right)\frac{\partial}{\partial \rho_*} \nonumber \\
I_1&=&\int_{-\infty}^{\infty} \bar{f}(\rho_*)Lf(\rho_*) \rho_*^2d\rho_*\nonumber\\
&=& \int_{-\infty}^0 \bar{f}(\rho_*)Lf(\rho_*) \rho_*^2d\rho_* + \int_0^{\infty} \bar{f}(\rho_*)Lf(\rho_*) \rho_*^2d\rho_* \nonumber \\
&=& \int_0^{\infty} \bar{f}(-\rho_*)Lf(-\rho_*) \rho_*^2d\rho_* + \int_0^{\infty} \bar{f}(\rho_*)Lf(\rho_*) \rho_*^2d\rho_* \nonumber \\
&=& I_{1,1} + I_{1,2} \\
\int_{S^2} I_{1,2}d\omega &=& \int_{S^2}\int_0^{\infty} \bar{f}(\rho_*)Lf(\rho_*) \rho_*^2d\rho_*d\omega \nonumber\\
&=& \int_{\Reals^3} \bar{f}(\rho_*)Lf(\rho_*) d^3x \nonumber
\end{eqnarray}
We now treat $\rho_*$ as a radial parameter in $\Reals^3$ and use $\frac{\partial}{\partial \rho_*} = \sum_{1=1}^3 \frac{x_i}{\rho_*}\frac{\partial}{\partial x_i}$
\begin{eqnarray}
\frac{-1}{2}L &=& \sum_{i=1}^3\left(\frac{h'(\rho_*)}{\rho_*}-\frac{x_i^2h'(\rho_*)}{\rho_*^3}+\frac{x_i^2h''(\rho_*)}{\rho_*^2} + \frac{x_ih'(\rho_*)}{\rho_*}\frac{\partial}{\partial x_i}\right)\frac{\partial}{\partial \rho_*} \nonumber\\
&=& \sum_{i=1}^3 \frac{\partial}{\partial x_i}\frac{x_ih'(\rho_*)}{\rho_*}\frac{\partial}{\partial \rho_*} \nonumber\\
L &=& -2\sum_{i,j=1}^3 \frac{\partial}{\partial x_i}\frac{x_ih'(\rho_*)}{\rho_*}\frac{x_j}{\rho_*}\frac{\partial}{\partial x_j} \label{I1Lformula}
\end{eqnarray}
Since $\frac{h'(\rho_*)}{\rho_*^2}$ is positive, from equation (\ref{I1Lformula}) we conclude that $L$ is a positive operator, and hence that $I_{1,2}, I_{1,1}$, and $I_1$ are all positive. 

We now return our attention to $I_2$. 
\begin{eqnarray}
2Mg'(r_*) &=& \frac{1}{(1+(\frac{\rho_*}{2M})^2)^{\sigma}} \nonumber\\
2Mg''(r_*) &=& \frac{-\sigma}{(1+(\frac{\rho_*}{2M})^2)^{\sigma+1}}2\frac{\rho_*}{2M}\frac{1}{2M} \nonumber\\
2Mg'''(r_*) &=& \frac{\sigma(\sigma+1)}{(1+(\frac{\rho_*}{2M})^2)^{\sigma+2}}\frac{4}{4M^2}(\frac{\rho_*}{2M})^2 +\frac{-\sigma}{(1+(\frac{\rho_*}{2M})^2)^{\sigma+1}}\left(\frac{2}{4M^2}\right) \nonumber\\
\frac{-2g''(r_*)}{\rho_*}-\frac{g'''(r_*)}{2} &=& \frac{\sigma}{(1+(\frac{\rho_*}{2M})^2)^{\sigma+2}}\frac{1}{(2M)^3}\left[4(1+(\frac{\rho_*}{2M})^2)-2(\sigma+1)(\frac{\rho_*}{2M})^2+(1+(\frac{\rho_*}{2M})^2)\right] \nonumber\\
&=& \frac{\sigma}{(1+(\frac{\rho_*}{2M})^2)^{\sigma+2}}\frac{1}{(2M)^3}\left[5+(3-2\sigma)(\frac{\rho_*}{2M})^2\right] 
\end{eqnarray}
We require $\sigma\leq \frac{3}{2}$ for this to be positive. 
\end{proof}

We go onto calculate the contribution from the potential $V$. Although this is negative in the region $r\in(\frac{8M}{3},3M)$, the lower bound provided in the previous result can be shown to dominate. Thus the contributions from the second radial derivative and from the potential combine to give a term that is bounded below by the local decay term we desire; however, the constants involved in this bound depend on $\sigma$ and vanish as $\sigma\rightarrow\frac{3}{2}$. In proving this result it is necessary to compare explicit values. 

\begin{lemma}
\label{MorawetzdrssVbound}
For $\sigma\in(\frac{1}{2},\frac{3}{2})$, $\exists c_{1,\sigma}>0:$
\begin{equation}
i[-\drss+V,\gamma] \geq \frac{c_{1,\sigma}}{(1+(\frac{\rho_*}{2M})^2)^{\sigma+1}}
\end{equation}
\end{lemma}
\begin{proof}
We first compute the commutator of the potential $V$ with $\gamma$.
\begin{eqnarray}
i[V,\gamma] &=& -g\drs V \nonumber\\
&=& -g\drs \left(\frac{2M}{r^3}\left(1-\frac{2M}{r}\right)\right) \nonumber\\
&=& -g\left( -3\frac{2M}{r^4} +4\frac{(2M)^2}{r^5}\right)\left(1-\frac{2M}{r}\right) \nonumber\\
&=& g\left(3-\frac{8M}{r}\right)\frac{2M}{r^4}\left(1-\frac{2M}{r}\right)
\end{eqnarray}
All these terms are positive except $3-\frac{8M}{r}$ and $g$ which go from negative to positive at $r=\frac{8M}{3}$ and $r=\alpha=3M$ respectively. Thus the commutator is positive except in the region $r\in(\frac{8M}{3},3M)$. We will show that in this region the commutator $i[-\drss,\gamma]$ will dominate. For the rest of this lemma, all calculations will be in the region $r\in(\frac{8M}{3},3M)$
\begin{eqnarray}
|r_*-\alpha_*| &\leq& \max_{r\in(\frac{8M}{3},3M)}\{\frac{dr_*}{dr}\} |r-\alpha| \nonumber\\
&\leq& \max_{r\in(\frac{8M}{3},3M)}\{\left(1-\frac{2M}{r}\right)^{-1}\} |r-\alpha| = 4|r-\alpha| \\
g(r_*) &\leq& \frac{|r_*-\alpha_*|}{2M} \nonumber\\
&\leq& 4|\frac{r}{2M}-\frac{3}{2}| \\
1-\frac{2M}{r} &\leq& \max_{r\in(\frac{8M}{3},3M)} \left(1-\frac{2M}{r}\right) = \frac{1}{3} \\
\frac{1}{r} &\leq& \frac{3}{8M} = \frac{1}{2M} \frac{3}{4} \\
|i[V,\gamma]| &\leq& |g| |3-\frac{8M}{r}| \frac{2M}{r^4} |1-\frac{2M}{r}| \nonumber\\
&\leq& 4|\frac{r}{2M}-\frac{3}{2}| |3-\frac{8M}{r}| \frac{1}{(2M)^3}(\frac{3}{4})^4\frac{1}{3} \nonumber\\
&\leq& |\frac{r}{2M}-\frac{3}{2}| |\frac{r}{2M}-\frac{4}{3}| \frac{3(2M)}{r} \frac{1}{(2M)^3}(\frac{3}{4})^3
\end{eqnarray}
To bound this note that $|(x-\frac{3}{2})(x-\frac{4}{3})|\leq\frac{1}{12^2}$ on $(\frac{4}{3},\frac{3}{2})$.
\begin{equation}
|i[V,\gamma]| \leq \frac{1}{12^2} \frac{1}{(2M)^3} 3 (\frac{3}{4})^3
\end{equation}
From (\ref{drssbound}), we have a formula for $i[-\drss,\gamma]$. From the bounds on $\sigma$ and $r$ we can bound this from below in the desired region. 
\begin{eqnarray}
i[-\drss,\gamma] &\geq& \frac{1}{(2M)^3}\frac{\sigma}{(1+(\frac{r_*-\alpha_*}{2M})^2)^{\sigma+2}}[5+(3-2\sigma)(\frac{r_*-\alpha_*}{2M})^2] \nonumber\\
&\geq& \frac{1}{(2M)^3}\frac{5\sigma}{(1+\frac{4}{9})^{\sigma+2}} \nonumber\\
&\geq& \frac{1}{(2M)^3} \frac{5}{2} (\frac{9}{13})^{\frac{7}{2}}\nonumber
\end{eqnarray}
Combining these results, and continuing to work in the region $r\in(\frac{8M}{3},3M)$, we find
\begin{eqnarray}
i[-\drss+V,\gamma] &\geq& \frac{1}{(2M)^3}\left(\frac{5}{2}\left(\frac{9}{13}\right)^{\frac{7}{2}} -\frac{1}{12^2}3\left(\frac{3}{4}\right)^3\right) \nonumber\\
&\geq& \frac{1}{(2M)^3} (.276-.009) \nonumber\\
&\geq& \frac{1}{(2M)^3} .25 \nonumber\\
&>&0
\end{eqnarray}
The strict positivity of this term in the region considered, the positivity of $i[V,\gamma]$ outside this region and (\ref{drssbound}) combine to give $c_{1,\sigma}$ such that
\begin{equation}
i[-\drss+V,\gamma] \geq \frac{c_{1,\sigma}}{(1+(\frac{\rho_*}{2M})^2)^{\sigma+1}}
\end{equation}
\end{proof}

The contribution from the non-linear term is calculated below and found to provide a localised non-linear estimate. Integration by parts provides two terms to integrate against: $r^2g'$, which is positive, and $2rg(1-\frac{2M}{r}(1+\frac{1}{p-1}))$ which is positive for all $r$ when $p=3$, but has regions in which it is negative for other values of $p$. This represents some form of competition between the non-linearity and the curvature which is exactly balanced at $p=3$. We expect that for $p$ close enough to $3$, the other term, $r^2g'$, contributes enough to provide positivity for all $r$. Numerical calculations verify this, but the exact range of $\sigma$ and $p$ for which this holds is difficult to estimate. 

\begin{lemma}
\label{MorawetzNonLinBound}
For $\sigma\in(\frac{1}{2},\frac{3}{2})$, $p=3$, and $u\in D(H_p,\gamma)$, $\exists c_{2,\sigma}>0:$
\begin{equation}
\langle u,i[\lambda r^{1-p}|u|^{p-1}(1-\frac{2M}{r}),\gamma]u\rangle \geq c_{2,\sigma} \lambda \int_\starman (1-\frac{2M}{r}) r^{-1-p}|u|^{p+1} d^{3}\mu
\end{equation}
\end{lemma}
\begin{proof}
\begin{equation}
\langle u,i[\lambda r^{1-p}|u|^{p-1}(1-\frac{2M}{r}),\gamma]u\rangle \pushleft
\end{equation}
\begin{eqnarray}
&=& -\int |u|^{2}g\drs(\lambda r^{1-p}|u|^{p-1}(1-\frac{2M}{r}))d^{3}\mu \nonumber\\
&=& -\int g \lambda r^2(1-\frac{2M}{r})^{\frac{-2}{p-1}}\frac{p-1}{p+1}\drs (r^{-1-p}|u^{p+1}(1-\frac{2M}{r})^{\frac{p+1}{p-1}}) d^{3}\mu \nonumber\\
&=& \lambda\frac{p-1}{p+1}\int \drs( g r^2 (1-\frac{2M}{r})^{\frac{-2}{p+1}}) r^{-1-p}|u|^{p+1}(1-\frac{2M}{r})^{\frac{p+1}{p-1}} d^{3}\mu \nonumber
\end{eqnarray}
To get a bound on this, we need to consider the derivative in the integral. 
\begin{eqnarray}
&&\drs(r^2 g (1-\frac{2M}{r})^{\frac{-2}{p-1}}) \nonumber\\
&&= (1-\frac{2M}{r})^{\frac{-2}{p-1}}\drs (r^2 g) -\frac{2}{p-1}r^2g(1-\frac{2M}{r})^{\frac{-p-3}{p-1}}\frac{2M}{r^2}(1-\frac{2M}{r}) \nonumber\\
&&= (1-\frac{2M}{r})^{\frac{-2}{p-1}}[\drs(r^2 g)-\frac{2}{p-1}2Mg] \nonumber\\
&&= (1-\frac{2M}{r})^{\frac{-2}{p-1}}[2r(1-\frac{2M}{r})g +r^2g' -\frac{2}{p-1}2Mg] \nonumber\\
&&= (1-\frac{2M}{r})^{\frac{-2}{p-1}}[2r(1-\frac{2M}{r}-\frac{2M}{r(p-1)})g +r^2g'] \nonumber\\
&&= (1-\frac{2M}{r})^{\frac{-2}{p-1}}[2rg(1-\frac{2M}{r}(1+\frac{1}{p-1})) +r^2g'] 
\end{eqnarray}
In the case that $p=3$, this becomes 
\begin{equation}
(1-\frac{2M}{r})^{\frac{-2}{p-1}}[2rg(1-\frac{3M}{r}) +r^2g'] 
\end{equation}
which is positive since the product $g(1-\frac{3M}{r})$ and all the other terms are positive. This gives the desired result. 
\end{proof}

The previous lemmas are all combined to provide the following relation between the commutator and localised norms. 

\begin{theorem}
\label{Morawetzthm}
If $p=3$, and $\sigma\in(\frac{1}{2},\frac{3}{2})$, and $u\in D(H_p,\gamma)$, then $\exists c_{1,\sigma}, c_{2,\sigma}$: 
\begin{eqnarray}
\langle u,i[H,\gamma_\sigma]u\rangle &\geq& c_{1,\sigma}\int_{S} \frac{|u|^2}{(1+(\frac{r_*-\alpha_*}{2M})^2)^{\sigma+1}} d^3\mu \label{LinMor}\\
\langle u,i[\lambda r^{1-p}|u|^{p-1}(1-\frac{2M}{r}),\gamma_{\sigma}]u\rangle &\geq& c_{2,\sigma} \lambda \int_{S} (1-\frac{2M}{r})r^{-1-p}|u|^{p+1} d^3\mu \label{NonLinMor}
\end{eqnarray}
\end{theorem}
\begin{proof}
Combine lemmas \ref{MorawetzNonLinBound}, \ref{MorawetzdrssVbound}, and \ref{MorawetzAngular}. 
\end{proof}

\Section{Local Decay}
\label{secLocalDecay}
We begin with a lemma on the derivative of the local $L^2$ norm. Because we are going to use a bootstrap argument, we need to find the consequences of a certain integral inequality, which are described in lemma \ref{TechnicalLemma}. We will then use the Heisenberg equation for the wave equation and the lower bound on the Morawetz type commutator to prove local decay. 

\begin{lemma}
For $u$ a solution to the non-linear wave equation in $\mathcal{H}$, $\sigma\in(\frac{1}{2},\frac{3}{2})$
\begin{equation}
| \dt (\|(1+(\frac{r_*-\alpha_*}{2M})^2)^{-\frac{\sigma+1}{2}}u\|) | \leq \|u\|_{\mathcal{H}}
\end{equation}
\end{lemma}
\begin{proof}
We define the localisation function $f$ and then perform computations similar to those for theorem \ref{L2Growth}.
\begin{eqnarray}
f &\defin& (1+(\frac{r_*-\alpha_*}{2M})^2)^{-\frac{\sigma+1}{2}} \nonumber\\
\dt \|fu\|^2 &=& \dt\langle fu,fu\rangle \nonumber\\
2\|fu\| \dt \|fu\| &=& 2\langle fu, f\dot{u}\rangle\nonumber\\
\|fu\| | \dt\|fu\| |&\leq& \|fu\| \|f\dot{u}\|\nonumber\\
| \dt\|fu\| |&\leq& \|f\dot{u}\| \leq \|\dot{u}\| \nonumber\\
&\leq& \|u\|_{\mathcal{H}}\nonumber
\end{eqnarray}
\end{proof}

The following lemma simply derives a decay property for a function satisfying a certain integral inequality. It is not specific to the problem at hand; however, we will use it to derive local decay by taking the localised norm as the function $\theta(t)$ in this lemma. 

\begin{lemma}
\label{TechnicalLemma}
Given $\theta:\Reals\rightarrow\Reals^+$ with uniformly bounded derivative. Suppose for some $\epsilon\in(0,\frac{1}{3}), \exists C_1, C_2, T: \forall t>T$
\begin{equation}
\int_0^t\theta(\tau)^2d\tau \leq C_1 + C_2t^{\epsilon}\theta(t)^{1-\epsilon} \label{TechLemmaIntegralCondition}
\end{equation}
then $\exists\{t_i\}: t_i\rightarrow\infty$, and $t_i^\epsilon\theta(t_i)^{1-\epsilon}\rightarrow 0$.
\end{lemma}
\begin{proof}
Successively stronger bounds on $\theta(t)$ will be proven. 

The bound on the derivative implies $\theta(t)$ is linearly bounded above. 

Suppose there is no sequence $\{t_i\}$ on which $\theta(t_i)\rightarrow 0$, then $\theta(t)$ is bounded below by a constant. By the integral condition $t^\epsilon\theta(t)^{1-\epsilon}$ is also bounded below by a linear function. Now $\theta(t)^2$ is bounded below by a quadratic and its integral is bounded by a cubic. This contradicts the linear upper bound of the integral. Therefore, there is a sequence $\{t_i\}$ such that $t_i\rightarrow \infty$ and $\theta(t_i)\rightarrow 0$. 

Since $\epsilon<\frac{1}{3}$, 
\hide{
\begin{eqnarray}
\frac{-\epsilon}{1-\epsilon}&>& -\frac{1}{2} \nonumber
\end{eqnarray}
and
\begin{eqnarray}
\frac{1-\epsilon}{1+\epsilon} &>& \frac{\epsilon}{1-\epsilon}\nonumber
\end{eqnarray}
Thus,} 
there exists $r<1$ such that
\begin{equation*}
-\frac{1-\epsilon}{2-r+r\epsilon} < -\frac{\epsilon}{1-\epsilon} 
\end{equation*}
Now choose $\delta$ negative such that $\delta \geq \frac{-\epsilon}{1-\epsilon} $, from which it follows that
\begin{eqnarray}
r\delta &<& 1+\frac{2\delta}{1-\epsilon}\label{rdeltabound}
\end{eqnarray}

Suppose $\exists K>0, S>0: \forall t>S: \theta(t)>Kt^{\delta}$, then by the integral condition (\ref{TechLemmaIntegralCondition}) there is a positive $C_3$ such that
\begin{equation}
C_3 t^{1+2\delta} \leq C_1 + C_2t^{\epsilon}\theta(t)^{1-\epsilon}
\end{equation}
Since $\delta\geq\frac{-\epsilon}{1-\epsilon}>\frac{-1}{2}$, $1+2\delta$ is strictly positive. Thus for sufficiently large $t$, there are constants $C_4, C_5$ such that
\begin{eqnarray}
C_4t^{1+2\delta} &\leq& t^{\epsilon}\theta(t)^{1-\epsilon} \nonumber\\
C_5t^{1+\frac{2\delta}{1-\epsilon}} &\leq& \theta(t) \nonumber
\end{eqnarray}
If $\delta$ is sufficiently close to zero, then this contradicts the previous result that $\theta(t)$ goes to zero on a subsequence. If $\delta$ is larger, then we use the fact that by equation (\ref{rdeltabound}), $\theta(t)\geq C_5t^{r \delta}$. This implies the original assumed lower bound is replaced by the larger lower bound $t^{r\delta}$. Repeated iterations of this process shows that $\theta(t)$ is bounded below by $t^{r^n\delta}$ for any $n$. Since $r<1$, this reduces the situation to the $\delta$ close to zero case, which led to a contradiction. Thus $\theta(t)$ can not be bounded below by a function of the form $Kt^{\delta}$ for $\delta\geq\frac{-\epsilon}{1-\epsilon}$. In particular, $\theta(t)$ can not be bounded below by a function of the form $Kt^{\frac{-\epsilon}{1-\epsilon}}$ and so the desired subsequence must exist.
\end{proof}

\begin{theorem}
If $u$ is a real valued solution of the non-linear wave equation in $D(H_p,\gamma)$, at $t=0$, $u=f$, $\dot{u}=g$, $\beta>\frac{3}{2}$, $p=3$, and $\sqrt{E}$ is the energy of $u$, then $\exists C_{1,\beta}, C_{2,\beta}$
\begin{eqnarray}
\int_0^T \|(1+(\frac{r_*-\alpha_*}{2M})^2)^{-\frac{\beta}{2}}u\|^2dt &\leq& C_{1,\beta} \sqrt{E} (\sqrt{E}+\|f\|_{L^2}) \\
\int_0^T \int_{\starman} \lambda(1-\frac{2M}{r}) r^{-1-p}|u|^{p+1} d^3\mu dt &\leq& C_{2,\beta} \sqrt{E}(\sqrt{E}+\|f\|_{L^2})
\end{eqnarray}
\end{theorem}
\begin{proof}
Throughout these calculations $\|u\|=\|u\|_{L^2}$. We begin by assuming that $\beta=\sigma+1$ for $\sigma\in(\frac{1}{2},1]$ and apply the results of theorem \ref{Morawetzthm} (equation \ref{LDApplyMorawetz}). We then apply the commutation result, theorem \ref{ComRelthm}, to get an exact derivative which we can integrate (equation (\ref{LDApplyCommutator}). After some calculations, we apply theorem \ref{P1Mm} to remove some terms (equation (\ref{removeterms})) and then the Schwarz inequality. We then apply conservation of energy and theorem \ref{P2Mm} (equation \ref{LDApplyEnergyConservation}). 
\begin{equation*}
c_{1,\sigma}\int_0^T \|(1+(\frac{r_*-\alpha_*}{2M})^2)^{-\frac{\sigma+1}{2}}u\|^2 + c_{2,\sigma}\int_{\starman} \lambda(1-\frac{2M}{r}) r^{-1-p}|u|^{p+1} d^3\mu dt
\end{equation*}
\begin{eqnarray}
&\leq& |\int_0^T\langle u,i[H_p,\gamma]u\rangle dt| \label{LDApplyMorawetz}\\
&\leq& |\int_0^T \dt[\langle u,i\gamma\dot{u}\rangle -\langle\dot{u},i\gamma u\rangle]dt| \label{LDApplyCommutator}\\
&\leq& |[\langle u,\gamma\dot{u}\rangle-\langle\dot{u},\gamma u\rangle]_0^T| \nonumber\\
&\leq& |[\dt\langle u,\gamma u\rangle-2\langle\dot{u},\gamma u\rangle]_0^T| \nonumber\\
&\leq& |[\langle\dot{u},\gamma u\rangle]_0^T| \nonumber\\
&\leq& |[ \|\dot{u}\| \|\gamma u\|]_0^T| \label{removeterms}\\
&\leq& | [\sqrt{E}(C_\sigma \sqrt{E} + D \|(1+(\frac{r_*-\alpha_*}{2M})^2)^{-\sigma}u\|)]_0^T| \label{LDApplyEnergyConservation}\\
&\leq& \sqrt{E} [C_\sigma \sqrt{E} + D\|(1+(\frac{r_*-\alpha_*}{2M})^2)^{-\sigma}u\|]_{t=0} + \sqrt{E} [C_\sigma \sqrt{E} + D\|(1+(\frac{r_*-\alpha_*}{2M})^2)^{-\sigma}u\|]_{t=T} \nonumber\\
&\leq& \sqrt{E} [2C_\sigma \sqrt{E} + D\|f\|] + D \sqrt{E}\|(1+(\frac{r_*-\alpha_*}{2M})^2)^{-\sigma}u\|_{t=T} \label{beforeHolderInLocalDecay}
\end{eqnarray}
Since $1>\sigma>\frac{1}{2}$, we can choose $q$ so that $\frac{1}{2\sigma}+\frac{1}{2}<q<\frac{3}{2}$. We now define $p$ to be the conjugate exponent to $q$, note a bound on $\frac{1}{p}$, and define $\kappa$ in terms of $p$. 
\begin{eqnarray}
\frac{1}{p} 
&>&1-\frac{2}{3} = \frac{1}{3} \nonumber\\
\kappa &\defin& \frac{2}{p} \nonumber\\
\frac{2-\kappa}{2}q &=& 1\nonumber\\
q\sigma &>&\frac{\sigma+1}{2}\nonumber
\end{eqnarray}
We now apply H\" older's inequality to the last term in (\ref{beforeHolderInLocalDecay}) using conjugate exponents $p$ and $q$. We then apply theorem \ref{L2Growth}. 
\begin{eqnarray}
\|(1+(\frac{r_*-\alpha_*}{2M})^2)^{-\sigma} u\|&=& \int_{\starman}\frac{|u|^{\kappa}|u|^{2-\kappa}}{(1+(\frac{r_*-\alpha_*}{2M})^2)^{2\sigma}} d^{3}\mu \nonumber\\
&\leq& \left(\int_{\starman}|u|^{\frac{p\kappa}{2}}d^{3}\mu\right)^{\frac{1}{p}}\left(\int_{\starman}\frac{|u|^{(2-\kappa)q}}{(1+(\frac{r_*-\alpha_*}{2M})^2)^{2\sigma q}}d^{3}\mu\right)^{\frac{1}{q}} \nonumber\\
&\leq& \| |u|^{\frac{p\kappa}{2}}\|^{\frac{1}{p}} \|\frac{|u|^{\frac{(2-\kappa)q}{2}}}{(1+(\frac{r_*-\alpha_*}{2M})^2)^{q\sigma}} \|^{\frac{1}{q}} \nonumber\\
&\leq& \|u\|^{\frac{1}{p}} \|\frac{u}{(1+(\frac{r_*-\alpha_*}{2M})^2)^{\frac{\sigma+1}{2}}}\|^{1-\frac{1}{p}} \nonumber\\
&\leq& (\sqrt{E}T+\|f\|)^{\frac{1}{p}} \|\frac{u}{(1+(\frac{r_*-\alpha_*}{2M})^2)^{\frac{\sigma+1}{2}}}\|^{1-\frac{1}{p}} \nonumber
\end{eqnarray}
For sufficiently large values of $T$ we get
\begin{equation}
\|(1+(\frac{r_*-\alpha_*}{2M})^2)^{-\sigma} u\| \leq F T^{\frac{1}{p}} \|\frac{u}{(1+(\frac{r_*-\alpha_*}{2M})^2)^{\frac{\sigma+1}{2}}}\|^{1-\frac{1}{p}} \nonumber
\end{equation}
Substituting this into (\ref{beforeHolderInLocalDecay}), we get
\begin{eqnarray}
\int_0^T \|(1+(\frac{r_*-\alpha_*}{2M})^2)^{-\frac{\sigma+1}{2}}u\|^2 dt 
&\leq&\int_0^T \|(1+(\frac{r_*-\alpha_*}{2M})^2)^{-\frac{\sigma+1}{2}}u\|^2 + \int_{\starman} \lambda (1-\frac{2M}{r})r^{-1-p}|u|^{p+1} d^{3}\mu dt\nonumber\\
&\leq& 2\sqrt{E} (2C_\sigma \sqrt{E} + D\|f\|) + D \sqrt{E} F T^{\frac{1}{p}}\|(1+(\frac{r_*-\alpha_*}{2M})^2)^{-\frac{\sigma+1}{2}}u\|^{1-\frac{1}{p}} \nonumber\\
\end{eqnarray}
We can now apply lemma \ref{TechnicalLemma} to conclude that on a subsequence 
\begin{equation}
t^{\frac{1}{p}} \|(1+(\frac{r_*-\alpha_*}{2M})^2)^{-\frac{\sigma+1}{2}}u\|^{1-\frac{1}{p}} \rightarrow 0\nonumber
\end{equation}

Since the integral being bounded is monotonically increasing in time, we can conclude 
\begin{equation}
\int_0^T \left( \|(1+(\frac{r_*-\alpha_*}{2M})^2)^{-\frac{\sigma+1}{2}}u\|^2
+ \int_{\tilde{S}} \lambda(1-\frac{2M}{r}) r^{-1-p}|u|^{p+1} d^{3}\mu\right) dt \leq \sqrt{E}( 2C_\sigma \sqrt{E} +D\|f\|) \nonumber
\end{equation}

This proves the result for $\beta\in(\frac{3}{2},2]$. Since $(1+(\frac{r_*-\alpha_*}{2M})^2)^{-\beta}$ is a decreasing function of $\beta$, the result holds for all $\beta>\frac{3}{2}$. 
\end{proof}

\end{document}